\documentclass[12pt,a4paper]{article}
\usepackage{amssymb,amsmath,stmaryrd}
\usepackage[mathscr]{eucal}
\usepackage{epsfig}

\newtheorem{proposition}{Proposition}

\newcounter{remark}

\newcounter{example}
\newenvironment{example}
{\begin{quote}\textsc{Example} \stepcounter{example} \arabic{example}:}
{\end{quote}}
\newenvironment{proof}{\medskip\noindent \bf Proof: \rm}{\hspace*{\fill}
$\blacksquare$ \newline \medskip}  

\newcommand{\Sipos}{\v{S}ipo\v{s}}
\newcommand{\svee}{\operatornamewithlimits{\varovee}}
\newcommand{\swedge}{\operatornamewithlimits{\varowedge}}
\newcommand{\sign}{\mathrm{sign}\,}
\def \cho {\mathcal{C}}
\def \symcho {\check{\mathcal{C}}}
\def \sug {\mathcal{S}}
\def \symsug {\check{\mathcal{S}}}

\def\1{{\mathchoice {\rm 1\mskip-4mu l} {\rm 1\mskip-4mu l}
{\rm 1\mskip-4.5mu l} {\rm 1\mskip-5mu l}}}
\def \0{{\mathbb{O}}}

\begin{document}

\title{The Symmetric Sugeno Integral}

\author{\underline{Michel GRABISCH}\thanks{On leave from
THALES, Corporate Research Laboratory,
Domaine de Corbeville, 91404 Orsay Cedex, France.}\\
LIP6, UPMC\\
4, Place Jussieu, 75252 Paris, France\\
Tel (+33) 1-44-27-88-65, Fax (+33) 1-44-27-70-00\\
\normalsize email \texttt{Michel.Grabisch@lip6.fr}}

\date{}

\maketitle

\begin{abstract}
We propose an extension of the Sugeno integral for negative numbers, in the
spirit of the symmetric extension of Choquet integral, also called \Sipos\
integral.    Our framework is purely ordinal, since the Sugeno integral has its
interest when the underlying structure is ordinal. We begin by defining
negative numbers on a linearly ordered set, and we endow this new structure
with a suitable algebra, very close to the ring of real numbers. In a second
step, we introduce the M\"obius transform on this new structure. Lastly, we
define the symmetric Sugeno integral, and show its similarity with the
symmetric Choquet integral. 
\end{abstract}


\section{Introduction}
In the field of fuzzy measure theory, the Sugeno integral \cite{sug74} has been
the first proposed to compute an average value of some function with respect to
a fuzzy measure, and early applications of fuzzy measures in multicriteria
evaluation have used this integral. Later, Murofushi and Sugeno proposed the
use of a functional defined by Choquet \cite{cho53} as a better definition of
an integral with respect to a fuzzy measure. In decision making, the pioneering
works of Schmeidler \cite{sch86,sch89} brought also into light the so-called
Choquet integral, as a generalization of expected utility.

This attention focused on the Choquet integral gave rise to a rapid progress,
both on a pure mathematical point of view and in decision making. As a
consequence, the Sugeno integral has remained a little bit in the background
until recently, where it has been (re)discovered that the Sugeno integral could
have some interest in qualitative decision making, and more generally, whenever
qualitative or ordinal information is used. This is due to the ordinal nature
of its definition, which uses only \emph{min} and \emph{max}. See e.g.
\cite{duprsa01} and \cite{mar00a} for recent results on Sugeno integral.

This paper brings new results for the Sugeno integral, essentially motivated by
qualitative decision making, although having their own mathematical
interest. The starting point is the following: the Choquet integral can be
defined in two different ways for functions taking negative values, named after
Denneberg as \emph{symmetric} and \emph{asymmetric} Choquet integrals
\cite{den94}. The symmetric integral, which integrates separately positive and
negative values, is of particular interest in decision making, since it
reflects well the symmetry in behaviour of human decision making (gains and
losses are treated separately and differently). Cumulative Prospect Theory
\cite{tvka92} precisely models this symmetric behaviour, and is based on the
symmetric Choquet integral (see \cite{grla99} for a comparative study of
symmetric and asymmetric models in decision making). However, until now, there
was no proposal for defining the Sugeno integral over functions taking negative
values.

The aim of this paper is precisely to fill this gap, i.e. to define a symmetric
Sugeno integral, in a general way. A fundamental requirement here is to keep
the ordinal nature of the Sugeno integral, and our construction should work for
any function and fuzzy measure valued on a (possibly finite) totally ordered
set (ordinal scale), where no other operation than supremum and infimum are
defined. Our construction is in three steps:
\begin{itemize}
\item definition of the concept of negative value for an ordinal scale, leading
to a symmetric ordinal scale. Then we endow it with suitable operations so as
to build a structure close to a ring, and to mimic real numbers with usual ring
operations. 
\item definition of an ordinal M\"obius transform. It is known that the
M\"obius transform is a powerful tool for the analysis and representation of
fuzzy measures, or even more general set functions: it is a basic ingredient
for Dempster-Shafer theory \cite{sha76}, upper and lower probabilities
\cite{chja89}, cooperative game theory (where it is called \emph{dividend}),
and also pseudo-Boolean functions \cite{haho87}. Also,
the Choquet integral has a particularly simple form using the M\"obius
transform. Since fuzzy measures become valued on an ordinal scale, one has to
redefine the M\"obius transform accordingly.
\item definition of a symmetric Sugeno integral, based on the preceding two
steps. 
\end{itemize}
We will make an emphasis on the third step, since the first and second steps
have been solved in a previous paper by the author \cite{gra01d}, and we will
recall here the main results, limited to what is necessary for defining the
symmetric Sugeno integral.

Preliminary works on symmetric Sugeno integral by the author can be found in
\cite{gra00}. 

\section{Basic concepts}
We provide in this section the necessary material for the sequel. Since our
results are established on ordered structures, and will be compared to the
corresponding ones in the numerical case, we present in a first part definition
in the (classical) numerical setting. Then, is a second part, we present the
definitions on ordered structures. 

Let us consider a finite set $N=\{1,\ldots,n\}$. A \emph{fuzzy measure} or
\emph{capacity} on $N$
is a set function $v:\mathcal{P}(N)\longrightarrow [0,1]$ such that
$v(\emptyset)=0$, $v(N) = 1$, and $v(A)\leq v(B)$ whenever $A\subset
B$.

The \emph{conjugate} fuzzy measure of $v$ is defined by $\bar{v}(A) = 1 -
v(A^c)$, where $A^c$ denotes the complement set of $A$. 

We present some examples of fuzzy measures which will be useful in the sequel. 
A \emph{unanimity game} $u_B$ is a fuzzy measure defined by:
\[
u_B(A) := \left\{	\begin{array}{ll}
			1, & \text{ if } A\supset B, A\neq \emptyset\\
			0, & \text{ otherwise},
			\end{array}	\right.
\]
for any $B\subset N$.  A \emph{possibility measure} $\Pi$ \cite{dupr85b} is a
fuzzy measure satisfying:
\[
\Pi(A\cup B) = \Pi(A)\vee\Pi(B), \quad \forall A,B\subset N.
\]
Due to this property, $\Pi$ can be defined unambiguously by giving its value on
singletons only. We write $\pi(i):=\Pi(\{i\})$, $i\in N$, and call $\pi$ the
\emph{possibility distribution} associated to or generating $\Pi$. 
 
A \emph{necessity measure} $N$ is the conjugate of a possibility measure. 

We introduce now integrals with respect to fuzzy measures. Let us consider a
function $f:N\longrightarrow \mathbb{R}^+$. We write for simplicity
$f_i:=f(i)$, for $i\in N$.  The \emph{Choquet integral} \cite{cho53} of $f$
w.r.t $v$ is defined by:
\begin{equation}
\label{eq:cho1}
\mathcal{C}_v(f) := \sum_{i=1}^n [f_{(i)} - f_{(i-1)}]v(A_{(i)}),
\end{equation}
where $\cdot_{(i)}$ indicates a permutation on $N$ so that
$f_{(1)}\leq f_{(2)}\leq\cdots\leq f_{(n)}$, and
$A_{(i)}:=\{(i),\ldots,(n)\}$. Also $f_{(0)}:=0$.

We introduce now the extension of Choquet integral for real-valued
functions. Let us consider $f:N\longrightarrow \mathbb{R}$, the two usual
definitions are given below:
\begin{align}
\symcho_v(f) & = \mathcal{C}_v(f^+) - \mathcal{C}_v(f^-) \label{eq:symcho}\\
\cho_v(f) & = \mathcal{C}_v(f^+) - \mathcal{C}_{\bar{v}}(f^-) \label{eq:asym}
\end{align}
where  $f^+, f^-$ are respectively
the positive and the negative parts of $f$, that is
$f^+=(f_1^+,\ldots,f_n^+)$, $f^-=(f_1^-,\ldots,f_n^-)$, $f_i^+ =
f_i\vee 0$, $f_i^- = -f_i\vee 0$. 

These extensions are respectively named symmetric and asymmetric
integrals by Denneberg \cite{den94}. The first one was in fact
proposed by \Sipos{} \cite{sip79}, and the second one is the usual
definition of the Choquet integral for real-valued functions (hence we keep the
same symbol). The terms symmetric
and asymmetric come from the following property:
\begin{align}
\mathcal{C}_v(-f) & = - \mathcal{C}_{\bar{v}}(f)\\ 
\check{\mathcal{C}}_v(-f) & = - \check{\mathcal{C}}_v(f), \label{eq:psym}
\end{align}
for any $f$ in $\mathbb{R}^n$. The explicit expression of $\symcho_v$ is:
\begin{align}
\symcho_{\mu}(f)  = & \sum_{i=1}^{p-1} (f_{(i)} - f_{(i+1)})
         \mu(\{(1),\ldots,(i)\})\nonumber \\
        & + f_{(p)}\mu(\{(1),\ldots,(p)\})  \nonumber \\
    & + f_{(p+1)}\mu(\{(p+1),\ldots,(n)\}) \nonumber \\
       & +\sum_{i=p+2}^n (f_{(i)} - f_{(i-1)})
         \mu(\{(i),\ldots,(n)\}) \label{eq:sip}
\end{align}
where $f_{(1)}\leq\cdots\leq f_{(p)} <0\leq f_{(p+1)}\leq\cdots\leq f_{(n)}$. 
\medskip

We turn now to the ordinal case. We consider a linearly ordered set $L^+$, with
bottom and top elements denoted by $\0$ and $\1$.  A \emph{negation} is a
mapping $n:L^+\longrightarrow L^+$ such that $n(n(x))=x$ and $x\leq y$ implies
$n(x)\geq n(y)$. Note that if $L^+$ is finite, then $n$ is unique and is simply
the reverse order of $L^+$.

A $L^+$-valued fuzzy measure
is a set function $v:\mathcal{P}(N)\longrightarrow L^+$, which assigns $\0$ to
the empty set and $\1$ to $N$, and satisfies monotonicity as above.  

The conjugation is defined thanks to the negation by $\overline{v}(A) =
n(v(A^c))$. The definition of unanimity game, possibility and
necessity measures are left unchanged (just replace 0,1 by $\0,\1$). 

The corresponding integral in the ordinal case is the Sugeno integral
\cite{sug74}. We consider  functions
$f:N\longrightarrow L^+$. The \emph{Sugeno integral} of $f$ with respect
to $v$ is defined by:
\begin{equation}
\sug_v(f) := \bigvee_{i=1}^n [f_{(i)}\wedge
v(\{(i),\ldots,(n)\})] 
\end{equation}
with same notations as above. 

\section{Symmetric ordered structures}
We consider a linearly ordered set $(L^+,\leq)$, with bottom and top
elements denoted by $\0$ and $\1$. We introduce $L^-:=\{-a|a\in L^+\}$, with
the reversed order, i.e. $-a\leq -b$ iff $a\geq b$ in $L^+$. The bottom and top
of $L^-$ are respectively $-\1$ and $-\0$. 

We denote by $L$ the  union of $L^+$ and $L^-$, with identification of
$-\0$ with $\0$. Top and bottom are respectively $\1$ and $-\1$. We call $L$ a
\emph{symmetric linearly ordered set}. 

We introduce some mappings on $L$. The \emph{reflection} maps $a\in L$ to $-a$,
and $-(-a)=a$ for any $a\in L$. We have:
\[
(-a)\vee(-b) = -(a\wedge b), \quad (-a)\wedge(-b) = -(a\vee b).
\]
The \emph{absolute value} of $a\in L$ is denoted $|a|$, and $|a|:=a$ if $a\in
L^+$, and $|a|=-a$ otherwise. The \emph{sign function} is defined by:
\[
\sign : L \rightarrow L     
\,,\quad \sign x=\left\{ \begin{array}{ll}
-{\1} & \mbox{ for } x < {\0}\\
{\0} & \mbox{ for } x = {\0}\\
{\1} & \mbox{ for } x > {\0}
\end{array}
\right. \,.
\]

Our aim is to endow $L$ with operations similar to usual operations $+,\cdot$
on $\mathbb{R}$, so that the algebraic structure is close to a ring. However,
since our aim is to extend the Sugeno integral which is based on minimum and
maximum, we require that the restriction to $L^+$ of these new operations are
precisely minimum and maximum. We call \emph{symmetric maximum} and
\emph{symmetric minimum} these new operations, which we denote $\svee$ and
$\swedge$ respectively. 

\begin{figure}[htb]
\begin{center}
$
\epsfysize=7cm
\epsfbox{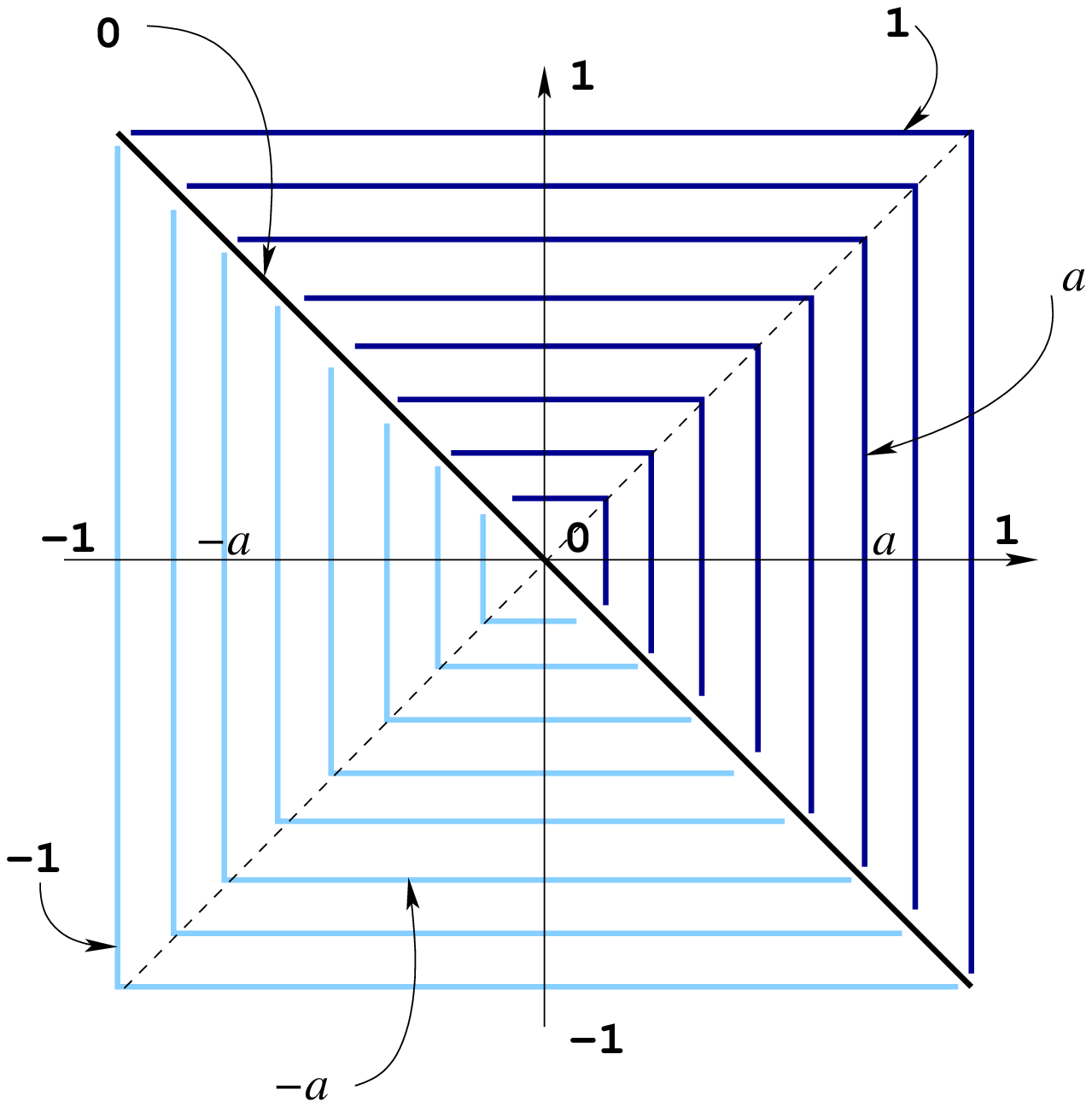}
$
\end{center}
\caption{Constant level curves of the symmetric maximum}
\label{fig:symmax}
\end{figure}

In \cite{gra01d}, it is shown that, based on the above requirements, the
``best'' possible definition (in the sense of being close to a ring) is given
as follows:
\begin{equation}
\label{eq:symmax}
a \svee b := \left\{ \begin{array}{ll}
-(|a| \vee |b|) & \mbox{ if } b \neq -a \mbox{ and either } |a| \vee |b|=-a \mbox{ or } =-b
\\
{\mathbb O} & \mbox{ if } b=-a \\
|a| \vee |b| & \mbox{ else.}
\end{array}
\right.
\end{equation}
Observe that, except for the case $b=-a$, $a \svee b$ equals the absolutely
larger one of the two elements $a$ and $b$. Figure \ref{fig:symmax} gives the constant level curves of this operation.

\begin{figure}[htb]
\begin{center}
$
\epsfysize=7cm
\epsfbox{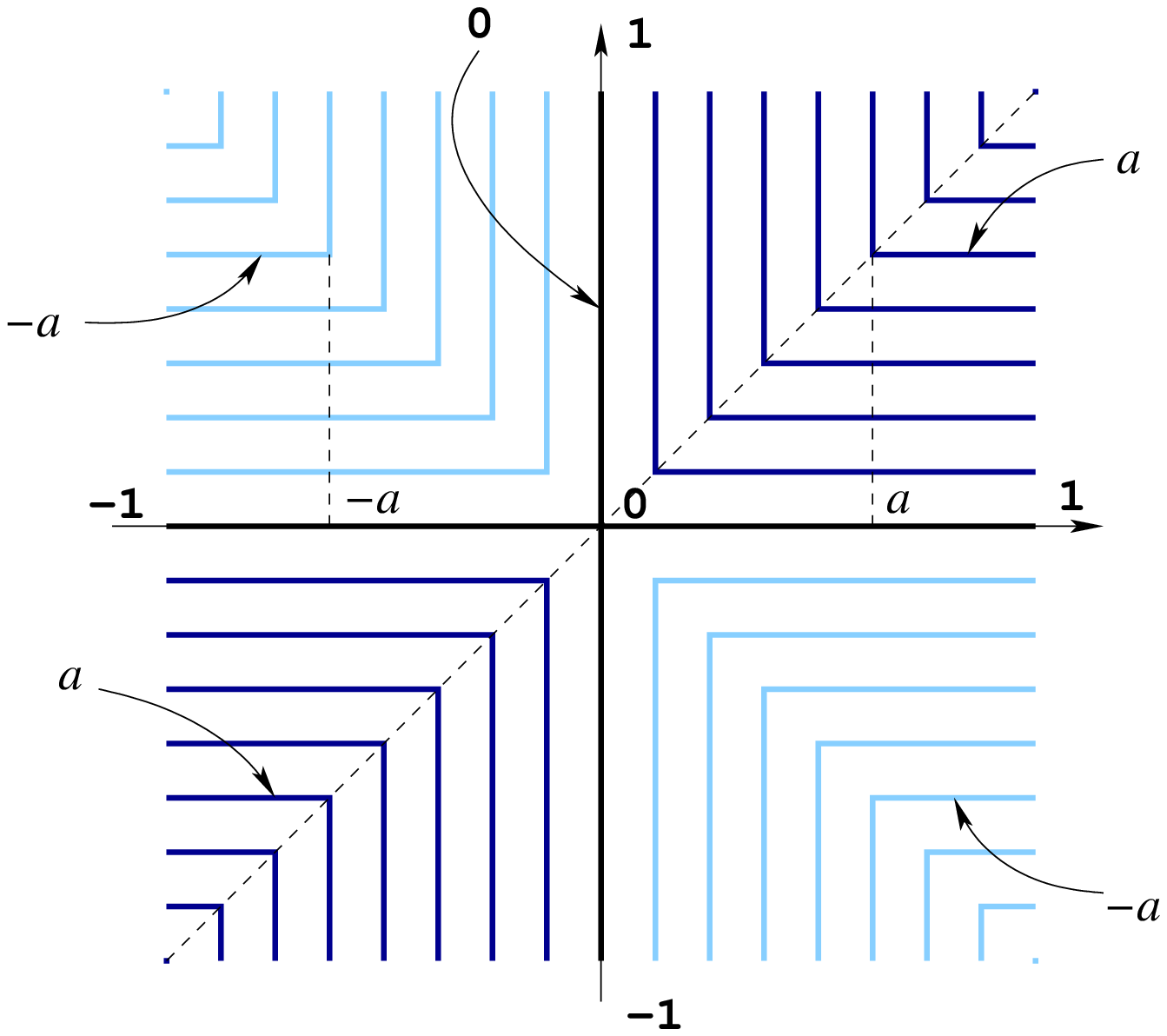}
$
\end{center}
\caption{Constant level curves of the symmetric minimum}
\label{fig:symmin}
\end{figure}

The symmetric minimum is defined as follows.
\begin{equation}
a \swedge b := \left\{ \begin{array}{ll}
-(|a| \wedge |b|) & \mbox{ if } \sign a \neq \sign b \\
|a| \wedge |b| & \mbox{ else.}
\end{array}
\right. 
\end{equation}
The absolute value of $a \swedge b$ equals $|a| \wedge |b|$ and $a \swedge b <
{\0}$ iff the two elements $a$ and $b$ have opposite signs. Figure
\ref{fig:symmin} shows the constant level curves of the symmetric minimum.
Another equivalent formulation of
these two operations is due to Marichal \cite{mar01}, when $L$ is a symmetric
real interval. It clearly shows the
relationship with the ring of real numbers.
\begin{align}
a\svee b & = \sign(a+b)(|a|\vee |b|)\\
a\swedge b & = \sign(a\cdot b)(|a|\wedge|b|).
\end{align}

The properties of $(L,\svee,\swedge)$, which is \emph{not} a ring, are
summarized below.  
\begin{proposition}
\label{prop:ring}
The structure $(L,\varovee,\varowedge)$ has the
following properties.
\begin{itemize}
\item [(i)] $\varovee$ is commutative.
\item [(ii)] $\0$ is the unique neutral element of $\varovee$, and the
unique absorbant element of $\varowedge$.
\item [(iii)] $a\varovee -a=\0$, for all $a\in L$.
\item [(iv)] $-(a\varovee b) = (-a)\varovee (-b)$.
\item [(v)] $\varovee$ is associative for any expression involving
$a_1,\ldots,a_n$, $a_i\in L$, such  that $\bigvee_{i=1}^n a_i\neq
-\bigwedge_{i=1}^n a_i$.
\item [(vi)] $\varowedge$ is commutative.
\item [(vii)] $\1$ is the unique neutral element of $\varowedge$, and the
unique absorbant
element of $\varovee$.
\item [(viii)] $\varowedge$ is associative on $L$.
\item [(ix)] $\varowedge$ is  distributive w.r.t $\varovee$ in
$L^+$ and $L^-$.
\end{itemize}
\end{proposition}
(v) shows that $\svee_{i=1}^n a_i$ is unambiguously defined iff
$\bigvee_{i=1}^n a_i\neq -\bigwedge_{i=1}^n a_i$. If equality occurs,
we can propose several \emph{rules of computation} which ensure
uniqueness, among which the following ones \cite{gra01d}:
\begin{enumerate}
\item Put $\svee_{i=1}^n a_i=0$. This corresponds to combine separately positive
 and negative values. We denote this rule by $\lfloor\svee_{i=1}^n a_i\rfloor$,
 and it is defined by:
\[
\lfloor\svee_{i=1}^n a_i\rfloor := \Big(\svee_{a_i\geq
\0}a_i\Big)\svee\Big(\svee_{a_i< \0}a_i\Big).
\]
\item Discard the pair(s) of opposite extremal values, successively until
  Condition (v) is satisfied. We denote this rule by $\lceil\svee_{i=1}^n
  a_i\rceil$, defined formally by:
\[
\lceil\svee_{a_i\in A}a_i\rceil := \svee_{a_i\in A\setminus \bar{A}}a_i,
\] 
 where $A:=a_1,\ldots,a_n$, and $\bar{A}:=\bar{a}_1,\ldots,\bar{a}_{2k}$ is the
 sequence of pairs of maximal opposite terms.
\item Discard as before pair(s) of maximal opposite terms, but with duplicates,
i.e. the set $\bar{A}$ contains in addition all duplicates of maximal opposite
terms.  This rule is denoted by $\langle\svee_{i=1}^n a_i\rangle$. 
\end{enumerate}
Taking for example $L=\mathbb{Z}$ and the sequence of numbers $3, 3, 3, 2, 1, 0,
-2, -3$, $-3$, for which associativity does not hold, the result for rule
$\lfloor\cdot\rfloor$ is 0, while we have:
\begin{align*}
\lceil 3\svee 3\svee 3 \svee 2\svee 1\svee 0\svee -2\svee -3\svee
-3\rceil = & 3\svee 2\svee 1\svee 0\svee -2 = 3\\
\langle 3\svee 3\svee 3 \svee 2\svee 1\svee 0\svee -2\svee -3\svee
-3\rangle = & 1 \svee 0 = 1. 
\end{align*}
We will use and comment on these rules in Section \ref{sec:alfo} (see also
\cite{gra01d} for a detailed study of their properties).  

The symmetric maximum with the $\lceil\cdot\rceil$ rule coincides with the limit
of some family of uni-norms proposed by Mesiar and Komornikov\'a \cite{meko98}.
We refer the reader to \cite{grbafo01} for details.

\section{The M\"obius transform on symmetric ordered structures}
Let us recall briefly some facts on the classical M\"obius transform (see
e.g. \cite{aig79,ber71}). Let $(X,\leq)$ be a locally finite (i.e. any
segment $[u,v]:=\{x\in X|u\leq x\leq v\}$ is finite) partially ordered
set (poset for short) possessing a unique
minimal element, denoted 0, and consider  $f,g$ two real-valued functions on $X$ such that
\begin{equation}
\label{eq:mob1}
g(x) = \sum_{y\leq x} f(y).
\end{equation}
A fundamental question in combinatorics is to solve this equation, i.e. to
recover $f$ from $g$. The solution is given through the \emph{M\"obius
function} $\mu(x,y)$ by
\begin{equation}
\label{eq:mob2}
f(x) = \sum_{y\leq x}\mu(y,x)g(y)
\end{equation}
where $\mu$ is defined inductively by
\[
\mu(x,y) = \left\{	\begin{array}{ll}
			1, & \text{ if } x=y\\
			-\sum_{x\leq t< y}\mu(x,t), & \text{ if } x< y\\
			0, &  \text{ otherwise}.
			\end{array}	\right.
\]

Taking for $X$ the Boolean lattice of subsets of a finite set
$N$, $f$ and $g$ are now set functions. In this case, for any $A\subset B\subset N$ we have $\mu(A,B)
= (-1)^{|B\setminus A|}$, and denoting set functions by $v,m$, formulas
(\ref{eq:mob1}) and (\ref{eq:mob2}) become
\begin{align}
v(A) = & \sum_{B\subset A} m(B) \label{eq:mob11}\\
m(A) = & \sum_{B\subset A}(-1)^{|A\setminus B|}v(B).\label{eq:mob21}
\end{align}
The set function $m$ is called the \emph{M\"obius transform} of $v$. When
necessary, we write $m^v$ to stress the fact it is the M\"obius transform of
$v$.

It is well known that the M\"obius transform is the coordinate vector of the
set function in the basis of unanimity games:
\begin{equation}
\label{eq:una}
v(A)  = \sum_{B\subset N} m^v(B)u_B(A), \quad \forall A\subset N.
\end{equation} 
Considering conjugate fuzzy measures $\overline{v}$, we can obtain a
decomposition with respect to $\overline{u_A}$ \cite{grla00a}:
\begin{equation}
\label{eq:conuna}
v(A) = \sum_{B\subset N} m^{\overline{v}}(B)\overline{u_B}(A), \quad \forall
A\subset N. 
\end{equation}
Noting that $\overline{u_B}(A)=1$ if $A\cap B\neq\emptyset$ and 0 otherwise, we
obtain
\begin{equation}
\label{eq:plau}
v(A) =\sum_{B|A\cap B\neq\emptyset} m^{\overline{v}}(B)= 1 - \sum_{B|A\cap B=\emptyset}m^{\overline{v}}(B). 
\end{equation}
The second expression comes from the fact that $v(N)= 1= \sum_{B\subset
N}m(B)$.  The first expression is well known in Dempster-Shafer theory
\cite{sha76}, which deals with fuzzy measures having non negative M\"obius
transforms, called \emph{belief functions}, whose conjugate are called
\emph{plausibility}.

Lastly, we recall the expressions of symmetric and asymmetric Choquet integrals
in terms of the M\"obius transform \cite{chja89,grla00a}.
\begin{align}
\mathcal{C}_v(f)  = & \sum_{A\subset N} m^v(A)\bigwedge_{i\in A} f_i \label{eq:chom}\\
\symcho_v(f)  = & \sum_{A\subset N} m^v(A)\left[ \bigwedge_{i\in A}
f_i^+ - \bigwedge_{i\in A} f_i^- \right] \label{eq:sipm} ,
\end{align}
for any $f\in \mathbb{R}^n$. 

\medskip

We turn now to our ordinal framework. Let $(L,\geq)$ be a symmetric linearly
ordered set. The \emph{successor} of any $x\in L$ is an element $y\in L$
such that $y>x$ and there is no $z$ such that $x<z<y$. We write $y\succ x$. We
consider two $L$-valued functions $f,g$ on $X$ satisfying the equation:
\begin{equation}
\label{eq:ordmob1}
g(x) = \svee_{y\leq x} f(y).
\end{equation}
Note that the above expression is well defined only if we use some rule of
computation, as the three rules proposed above. The study of the existence of
solutions to this equation is a difficult topic, partly solved in
\cite{gra01d}. We just mention here that if $|g|$ is isotone (i.e. $x\leq y$
implies $|g(x)|\leq |g(y)|$) and if either the
$\lfloor\cdot\rfloor$ or the $\langle\cdot\rangle$ rule is used, then there
exists many solutions, among which the canonical one, which is defined as
follows:  the \emph{canonical ordinal} M\"obius function
is defined by:
\begin{equation}
\mu(x,y):=\left\{	\begin{array}{ll}
			\1, & \text{ if } x=y\\
			-\1, & \text{ if } x\prec y\\
			\0, & \text{ otherwise.} 
			\end{array} \right.
\end{equation}
which leads to the \emph{canonical ordinal M\"obius transform} of $g$, defined by:
\[
 m^g(x) :=  g(x)\svee\Big[-\svee_{y\prec x}g(y)\Big],
\]
where in this expression the same computation rule used in (\ref{eq:ordmob1})
has to be applied.

This result is no more true for the $\lceil\cdot\rceil$ rule, which has no
solution in many cases.

It is possible to get the whole set of non negative solutions when $f,g$ are
valued on $L^+$ (so that computation rules become useless), and $g$ is isotone
(fuzzy measures correspond to this case, hence its interest).
\begin{proposition}
\label{prop:ordmob} \cite{gra01d}
For any non negative isotone function $g$,  any non negative solution of the
equation (\ref{eq:ordmob1}) lays in the interval $[m_*,m^*]$, defined by:
\begin{align*}
m^*(x) = & g(x), \quad\forall x\in X\\
m_*(x) = m^g(x) = & \left\{	\begin{array}{ll}
			g(x), & \text{ if } g(x)>g(y), \quad \forall y\prec x\\
			\0, & \text{ otherwise}
			\end{array}
		\right., \quad\forall x\in X.
\end{align*} 
\end{proposition}
Note that negative solutions exist. It is easy to see that 
\[
m_*(x) =  \left\{	\begin{array}{ll}
			g(x), & \text{ if } g(x)>g(y), \quad \forall y\prec
			x\\
			\text{any }e\in L,e\succ -g(x),  & \text{ otherwise}
			\end{array}
		\right.,
\]
$\forall x\in X$ is the least solution. However, negative solutions do not
possess good properties. It is to be noted that $m_*$ has been first proposed
as the (ordinal) M\"obius transform of a fuzzy measure by Marichal
\cite{mamato96} and Mesiar \cite{mes97} independently. See also preliminary
works of the author in \cite{gra97a}, and related works by De Baets
\cite{cade00}.

If there is no fear of ambiguity, we denote simply $m^g$ by $m$.
 Moreover, since our framework is ordinal in the rest of the paper,
we will omit to call it ``ordinal'', and will use the term ``classical''
M\"obius transform when referring to the usual definition. We denote by $[m]$
the interval $[m_*, m^*]$, and with some abuse of notation, any function in
this interval.

\medskip

From now on, we restrict to the case of fuzzy measures on a finite set
$N=\{1,\ldots,n\}$.

The M\"obius transform possesses many interesting properties, some of which are
listed below. 
\begin{itemize}
\item [(i)] the M\"obius transform can be written in a way which is very
similar to  formula (\ref{eq:mob21}):
\begin{equation}
m(A) := \bigvee_{B\subset A, |A\setminus B|
\mbox{\scriptsize  even}}v(B)
\svee\left(- \bigvee_{B\subset A, |A\setminus B| \mbox{\scriptsize odd}}v(B)\right)
\end{equation}
for any $A\subset N$ (see \cite{gra97a}).
\item [(ii)] the M\"obius transform is still the coordinate vector of any fuzzy
measure in the basis of unanimity games, but with a different decomposition:
\begin{equation}
\label{eq:orduna}
v(A) = \bigvee_{B\subset N} \Big([m](B)\wedge u_B(A)\Big), \quad \forall
A\subset N,
\end{equation}
where $[m]$ stands for any function in the interval $[m_*,m^*]$. But since the
decomposition is not unique, we have no more a basis.
\item [(iii)] Let $\pi$ a possibility distribution defined on $N$, such that
$\0<\pi(1)<\cdots<\pi(n)=\1$, and consider the associated possibility and
necessity measures $\Pi,\mathrm{N}$. Then their M\"obius transforms are given
by:
\[
m^\Pi(A) = \left\{	\begin{array}{ll}
			\Pi(\{i\}), & \text{ if } A=\{i\}, i\in N\\
			\0, & \text{ otherwise.}
			\end{array} \right.
\]
\[
m^{\mathrm{N}}(A) = \left\{	\begin{array}{ll}
			n(\Pi(\{i\}), & \text{ if }
			A=\{i+1,\ldots,n\}, i\in N\\
			\0, & \text{ otherwise.} 
			\end{array} \right.
\]
If for some $i$, we have $\pi(i)=\pi(i+1)$, then the result is unchanged for
$m^\Pi$, and for $m^{\mathrm{N}}$, we have $m^{\mathrm{N}}(\{i+1,\ldots,n\}) =
\0$. 

See \cite{gra01d} for a more general result, giving the M\"obius transform of a
conjugate fuzzy measure. 

Note that the ``focal elements'' (i.e., in the terminology of Shafer, the
subsets where the M\"obius transform is non zero) are singletons in the case of
the possibility measure, and are nested subsets for the necessity measure. This
result needs some comments. In the classical case, the M\"obius transform is
non zero only on singletons if and only if the fuzzy measure is a probability
measure, i.e. an additive measure. In our algebra based on min and max, the
corresponding notion is a ``maxitive'' measure, in other words, a possibility
measure. This shows the consistency of the construction. Now, it is also known
that the M\"obius transform of a necessity measure, in the classical case, is
non zero only for a chain (i.e. a set of nested subsets). It is very surprising
to get the same result here, and moreover, the chains are identical. Here there
is a discrepancy with the classical case, since probability measures are
self-conjugate, and possibility measures are not.
\item [(iv)] it is possible to derive a decomposition of a fuzzy measure
in terms of the conjugate of unanimity games, as in
(\ref{eq:conuna}). Specifically, using (\ref{eq:orduna}):
\begin{align*}
\overline{v}(A) = & n\Big(\svee_{B\subset N}([m^v](B)\wedge
u_B(A^c))\Big) \\
 = & n\Big(\svee_{B\subset N}([m^v](B)\wedge
n(\overline{u_B}(A)))\Big) \\
= & \swedge_{B\subset N}\Big[n([m^v](B))\vee\overline{u_B}(A)\Big]
\end{align*} 
since $n(a\vee b) = n(a)\wedge n(b)$ and $n(a\wedge b) = n(a)\vee n(b)$, for
any $a,b\in L^+$. Hence we get:
\begin{equation}
v(A) = \swedge_{B\subset N}\Big[n([m^{\overline{v}}](B))\vee\overline{u_B}(A)\Big].
\end{equation}
It can be shown \cite{gra01d} that $n([m^{\overline{v}}](B))=m^v(B^c)$
when $[m^v]\equiv v$. This shows that 
\[
v(A) = \swedge_{B\subset N}\Big[v(B^c)\vee\overline{u_B}(A)\Big].
\]
Now observe that $\overline{u_B}(A)=\1$ if $A\cap B\neq\emptyset$, and $\0$
otherwise. Thus,
\begin{equation}
v(A) = n\Big[\svee_{B\cap A=\emptyset}[m^{\overline{v}}](B)\Big]
\end{equation}
which is the exact counterpart of (\ref{eq:plau}).
\item [(v)] the author has proposed some time ago the notion of $k$-additive
measure \cite{gra96f}, i.e. a fuzzy measure whose (classical) M\"obius
transform vanishes for subsets of more than $k$ elements. As remarked by Mesiar
\cite{mes97}, the concept can be extended to the ordinal M\"obius
transform. The author called this \emph{$k$-possibility measures} ($k$-maxitive
measure in the terminology of Mesiar), since this
defines possibility distributions on subsets of at most $k$ elements (see
\cite{gra97a} for some properties of $k$-possibility measures). 
\end{itemize}
Finally, we indicate that, contrary to the classical case, the M\"obius
transform is not a ``linear'' operator on the set of fuzzy measures, where of
course ``linear'' is to be taken in the sense of ``maxitive''. This is shown by
the following example:
\begin{example}
Let us take $X$ to be the Boolean lattice $2^2$ whose elements are denoted
$\emptyset,\{1\},\{2\},\{1,2\}$, and consider two functions $g_1,g_2$ defined
as follows:
\begin{center}
\begin{tabular}{|c|c|c|c|c|}\hline
	& $\emptyset$ & $\{1\}$ &  $\{2\}$ & $\{1,2\}$ \\ \hline
$g_1$	& $\0$ & $\0$ &$\0$ &$\1$ \\ \hline
$g_2$	& $\0$ & $\1$ &$\1$ &$\1$ \\ \hline 
\end{tabular}
\end{center}
The computation of the M\"obius transform $m_*$ gives
\begin{center}
\begin{tabular}{|c|c|c|c|c|}\hline
	& $\emptyset$ & $\{1\}$ &  $\{2\}$ & $\{1,2\}$ \\ \hline
$m_*[g_1]$	& $\0$ & $\0$ &$\0$ &$\1$ \\ \hline
$m_*[g_2]$	& $\0$ & $\1$ &$\1$ &$\0$ \\ \hline 
\end{tabular}
\end{center}
Clearly, $g_1\svee g_2 = g_2$, but $m_*^{g_1}\svee m_*^{g_2} \neq
m_*^{g_2}$. 
\end{example}

\section{The symmetric Sugeno integral}
\label{sec:symsug}
It is possible to express the Sugeno integral with respect to the M\"obius
transform. Indeed, the following can be shown.
\begin{proposition}
\label{prop:sugmob}
For any function $f:N\longrightarrow L^+$ and any $L^+$-valued fuzzy measure
$v$ on $N$, the Sugeno integral of $f$ w.r.t. $v$ can be written as:
\[
\sug_v(f) := \bigvee_{A\subset N}\Bigg(\bigwedge_{i\in A}f_i\wedge [m](A)\Bigg)
\]
where $[m]$ is any function in $[m_*,m^*]$.
\end{proposition}
\begin{proof}
It suffices to prove that the relation holds for $m^*$ (i)  and
$m_*$ (ii). 

(i) We have, using distributivity of $\wedge,\vee$ and
monotonicity of $v$:
\begin{eqnarray*}
\bigvee_{A\subset N}\left(\bigwedge_{i\in A}f_i \wedge v(A)\right) & = &
\bigvee_{\substack{A\subset N\\A\ni (1)}}\left(f_{(1)}\wedge v(A)\right) \vee
\bigvee_{\substack{A\subset N\setminus(1)\\ A\ni
(2)}}\left(f_{(2)}\wedge v(A)\right)\vee \\ & & \cdots \vee (f_{(n)}\wedge
v(\{(n)\}) \\ & = & \Big(f_{(1)}\wedge \bigvee_{\substack{A\subset N\\A\ni
(1)}}v(A)\Big)\vee \Big(f_{(2)}\wedge \bigvee_{\substack{A\subset
N\setminus(1)\\ A\ni (2)}}v(A)\Big)\vee\\
& & \cdots \vee (f_{(n)}\wedge v(\{(n)\}) \\
& = & (f_{(1)}\wedge v(N)) \vee (f_{(2)}\wedge v(N\setminus (1))
\vee\cdots\\
& & \vee(f_{(n)}\wedge v(\{(n)\})\\
& = & \bigvee_{i=1}^n\left(f_{(i)}\wedge
v(\{(i),\ldots,(n)\})\right).
\end{eqnarray*}
(ii) For a given non empty $A\subset N$, if it exists some $j\in A$ such that
$v(A) = v(A\setminus j)$, then $\displaystyle v(A)\wedge\bigwedge_{i\in A}f_i
\leq v(A\setminus j)\wedge \bigwedge_{i\in A\setminus j}f_i$, hence the
corresponding term in the supremum over $N$ (in the expression with $m^*$) can
be deleted, or equivalently, $v(A)$ can be replaced by $\0$. But $m_*(A)=\0$
if $v(A) = v(A\setminus j)$ for some $j$, hence the result.
\end{proof}

The result with $m^*$ is already in the original work of Sugeno
\cite{sug74}. Also, Marichal has shown the above proposition using min-max
Boolean functions \cite{mar00a}. Note the analogy with the expression of the
Choquet integral using the M\"obius transform (see (\ref{eq:chom})).

We address now the problem of extending the definition of Sugeno integral for
functions which are $L$-valued, i.e. they may take ``negative''
values. We focus on the symmetric definition. 
Following what is done in the numerical case for the Choquet integral,
we propose the following definition for the symmetric Sugeno integral:
\begin{equation}
\label{eq:symsug}
\symsug_v(f) = \sug_v(f^+)\svee(-\sug_v(f^-))
\end{equation}
where $f^+:=f\vee \0$, $f^-:=(-f)\vee \0$. From the definition, it is immediate
that:
\begin{equation}
\label{eq:sym}
\symsug_v(-f) =   - \symsug_v(f)
\end{equation}
which justifies the name ``symmetric''. 

Let us express the symmetric integral in an explicit form, using the fuzzy
measure and its M\"obius transform.
\begin{proposition}
For any $f$ valued in $L$ and any fuzzy measure $v$ on $N$, 
\begin{align}
\label{eq:sugcap}
\symsug_v(f) := & \left[\svee_{i=1}^p \left(f_{(i)}\swedge
v(\{(1),\ldots,(i)\})\right)\right] \svee\nonumber \\
	&  \left[\svee_{i=p+1}^n \left(f_{(i)}\swedge
v(\{(i),\ldots,(n)\})\right)\right],
\end{align}
where $-\1\leq f_{(1)}\leq\cdots\leq f_{(p)}<\0$, and $\0\leq
f_{(p+1)}\leq\cdots\leq f_{(n)}\leq \1$. 
\begin{align}
\label{eq:sugmob}
\symsug_v(f) = & \left[\svee_{A\subset N^+}\left(m(A) \swedge\Bigg[\bigwedge_{i\in
A}f_i^+ \svee\Big(- \bigwedge_{i\in A}f_i^-\Big) \Bigg] \right)\right]\svee\nonumber\\
 & \left[\svee_{A\subset N^-}\left(m(A) \swedge\Bigg[\bigwedge_{i\in
A}f_i^+ \svee\Big(- \bigwedge_{i\in A}f_i^-\Big) \Bigg] \right)\right]\svee\nonumber\\
 & \left[\svee_{A^+,A^-\neq\emptyset}\left(m(A) \swedge\Bigg[\bigwedge_{i\in
A}f_i^+ \svee\Big(- \bigwedge_{i\in A}f_i^-\Big) \Bigg] \right)\right]
\end{align}
where $N^+:=\{i\in N|f_i\geq\0\}$, $N^-:=N\setminus
N^+$, $A^+:=A\cap N^+$, and $A^-:=A\cap N^-$. 
\end{proposition}
\begin{proof}
Let us show the first formula. By Prop. \ref{prop:sugmob}, we have:
\[
\Big[\bigvee_{A\subset N}(m(A)\wedge\bigwedge_{i\in
A}f_i^+)\Big] = \bigvee_{i=p+1}^n[f_{(i)}\wedge v(\{(i),\ldots,(n)\})]
\]
since $f_{(i)}=\0$ for $i<p$. Similarly, since $f^-_{(1)}\geq \cdots\geq
f^-_{(p)}$, we get:
\begin{align*}
- \bigvee_{A\subset N}\Big(m(A)\wedge
\bigwedge_{i\in A}f_i^-\Big) & = - \bigvee_{i=1}^p
\Big(f_{(i)}^-\wedge v(\{(1),\ldots,(i)\})\Big) \\ 
& = \svee_{i=1}^p \Big(-(f^-_{(i)}\wedge v(\{(1),\ldots,(i)\}))\Big)\\
& = \svee_{i=1}^p \Big((-f^-_{(i)})\swedge v(\{(1),\ldots,(i)\})\Big)\\
& = \svee_{i=1}^p \Big(f_{(i)}\swedge v(\{(1),\ldots,(i)\})\Big).
\end{align*}
Since the symmetric Sugeno integral is the ``sum'' of these two terms, the
result is proven.

Let us show the second formula.  By definition of $\symsug_v$ and
Prop. \ref{prop:sugmob}, we have:
\begin{align*}
\symsug_v(f) = & \Bigg[\bigvee_{A\subset N}\Big(m(A)\wedge\bigwedge_{i\in
A}f_i^+\Big)\Bigg] \svee\Bigg[-\bigvee_{A\subset
N}\Big(m(A)\wedge\bigwedge_{i\in A}f_i^-\Big)\Bigg]\\
 = & \Bigg[\bigvee_{A\subset N^+}\Big(m(A)\wedge\bigwedge_{i\in
A}f_i^+\Big)\Bigg] \svee\Bigg[-\bigvee_{A\subset
N^-}\Big(m(A)\wedge\bigwedge_{i\in A}f_i^-\Big)\Bigg]\\
 = & \Bigg[\svee_{A\subset N^+}\Bigg(m(A)\swedge\Big[\bigwedge_{i\in
A}f_i^+\svee\Big(- \bigwedge_{i\in
A}f_i^-\Big)\Big]\Bigg)\Bigg]\svee\\
 & \Bigg[\svee_{A\subset N^-}\Bigg(m(A)\swedge\Big[\bigwedge_{i\in
A}f_i^+\svee\Big(- \bigwedge_{i\in
A}f_i^-\Big)\Big]\Bigg)\Bigg].
\end{align*}
Observe that 
\[
\svee_{A^+,A^-\neq\emptyset}\left[m(A)\swedge\Bigg(\bigwedge_{i\in 
A}f_i^+\svee\Big(-\bigwedge_{i\in A} f_i^-\Big)\Bigg)\right] =
\0
\]
hence this last term can be added without changing the result.
\end{proof}
Let us make some comments on these results.
\begin{itemize}
\item [(i)] Both formulas are unambiguous with respect to possible
associativity problem, since positive terms and negative terms are separately
combined. 
\item [(ii)] Formula (\ref{eq:sugcap}) is very similar to (\ref{eq:sip}), which is the
expression of the symmetric Choquet integral.
\item [(iii)] Formula (\ref{eq:sugmob}) has in fact no computation interest,
since it is more complicated than necessary (see proof). Its interest lies in
the fact that it is a formula which is very close to the corresponding one
(\ref{eq:sipm}) for the symmetric Choquet integral.  Indeed, a summation over
$A\subset N$ can be partitioned into $A\subset N^+$, $A\subset N^-$, and
$A^+,A^-\neq \emptyset$. However, due to the lack of associativity, we cannot
write, as it was wrongly claimed in \cite{gra00}, that
\[
\symsug_v(f) = \svee_{A\subset N}\left(m(A) \swedge\left[\bigwedge_{i\in
A}f_i^+ \svee \Bigg(- \bigwedge_{i\in A}f_i^-\Bigg) \right] \right).
\]
\end{itemize}

\section{Alternative definitions for the symmetric Sugeno integral}
\label{sec:alfo}
While our definition of symmetric Sugeno integral (Eq. (\ref{eq:symsug})) seems
to be natural with respect to what is done for the Choquet integral, we may
think of other definitions, provided the symmetry property (\ref{eq:sym}) is
preserved. We propose the following ones, which satisfy the symmetry
requirement. 
\begin{align}
\label{eq:symsug1}
\symsug^1_v(f) = & \langle\svee_{A\subset N}\left(m(A)
\swedge\left[\bigwedge_{i\in A}f_i^+ \svee\Bigg(- \bigwedge_{i\in A}f_i^-\Bigg)
\right] 
\right)\rangle\\ \label{eq:symsug2}
\symsug^2_v(f) = & \langle\left[\svee_{i=1}^p \left(f_{(i)}\swedge
v(\{(1),\ldots,(i)\})\right)\right] \svee\nonumber \\
	&  \left[\svee_{i=p+1}^n \left(f_{(i)}\swedge
v(\{(i),\ldots,(n)\})\right)\right]\rangle.
\end{align}
The first one is suggested by Remark (iii) in Section \ref{sec:symsug}, in
order to avoid a complicated expression with the M\"obius transform. The second
one puts all terms of (\ref{eq:sugcap}) together and apply the rule of
computation. The following example shows that these formulas
and the original one are indeed different. 
\begin{example} Let us take $N=\{1,2,3\}$, $L=[-1,+1]$, $v$ and $f$
defined in the following tables.
\begin{center}
\begin{tabular}{|l|l|}\hline
$v(\{1\})= 0.3$ & $v(\{1,2\}) = 0.4$ \\ 
$v(\{2\})= 0.25$ & $v(\{1,3\}) = 0.3$ \\ 
$v(\{3\})= 0.2$ & $v(\{2,3\}) = 0.6$ \\ \hline
\end{tabular}
\hfill 
\begin{tabular}{|l|}\hline
$f(1) = -1$\\
$f(2) = 0.3$\\
$f(3) = 1$\\ \hline
\end{tabular}
\end{center}
Observe that the M\"obius transform $[m]$ is reduced to $v$, except for subset
$\{1,3\}$, where $[m](\{1,3\}) = [0,0.3]$. Let us compute $\sug_v(f^+)$ and
$\sug_v(f^-)$. We have:
\begin{align*}
\sug_v(f^+) = & (0.3\wedge 0.6)\vee(1\wedge 0.2) = 0.3\vee 0.2=0.3\\
\sug_v(f^-) = & (1\wedge 0.3) = 0.3.
\end{align*}
Hence, according to original definition (eq. (\ref{eq:symsug})), we get
$\symsug_v(f) = 0$. If we compute from (\ref{eq:symsug2}), we obtain:
\begin{align*}
\symsug^2_v(f) & = \langle  (-1\swedge 0.3)\svee(0.3\wedge 0.6)\svee(1\wedge
0.2)\rangle\\
&  = \langle -0.3\svee 0.3\svee 0.2\rangle = 0.2.
\end{align*}
Now, if we compute from (\ref{eq:symsug1}), we obtain:
\begin{align*}
\symsug^1_v(f) & = \langle (0.25\wedge 0.3)\svee(0.2\wedge 1)\svee(0.6\wedge
0.3)\svee(-1\swedge 0.3)\rangle\\
&  = \langle 0.25\svee 0.2\svee 0.3\svee -0.3\rangle = 0.25.
\end{align*}
Hence, all expressions lead to different results.
\end{example}

Let us comment about these formulas.
\begin{itemize}
\item  Clearly, $\symsug^1_v$ is not the
expression of $\symsug^2_v$ with the M\"obius transform, as one could
have expected, and it remains difficult to interpret.
\item Comparing the original formula and (\ref{eq:symsug2}) is easier since
they can be viewed as the same expression using different rules of computation,
which are $\langle\quad\rangle$ and $\lfloor\quad\rfloor$. This suggests a
third variant, using the rule denoted by $\lceil\quad\rceil$:
\begin{align}
\label{eq:symsug3}
\symsug^3_v(f) = & \lceil\left[\svee_{i=1}^p \left(f_{(i)}\swedge
v(\{(1),\ldots,(i)\})\right)\right] \svee\nonumber \\
	&  \left[\svee_{i=p+1}^n \left(f_{(i)}\swedge
v(\{(i),\ldots,(n)\})\right)\right]\rceil.
\end{align} 
$\symsug_v$ takes separately the maximum of positive and negative values, and
then compare the results. We obtain 0 as soon as the best positive value
equals in absolute value the worst negative one. $\symsug^2_v$ is more
discriminating than $\symsug_v$ in the sense that many cases where $\symsug_v$
gives 0 are distinguished by $\symsug^2_v$, since maximal opposite values are
discarded until they become different, in a way which is similar to the
``discrimin'' proposed by Dubois \emph{et al.} \cite{dufapr97}. This is also
the case for $\symsug^3_v$, except that multiple occurrences are not
removed. 
\item $\symsug^2_v$ is not monotonic in the sense that if
$f\geq f'$, it may happen that $\symsug^2_v(f)<\symsug^2_v(f')$, since the rule
$\langle\rangle$ is not monotonic, as shown by the following example:
\begin{example}
Let us consider the following values $a_i,b_i\in \mathbb{R}$, $i=1,\ldots,5$. 
\begin{center}
\begin{tabular}{|c|ccccc|}\hline
$i$ & 1 & 2 & 3 & 4 & 5 \\ \hline
$a_i$ & -5 & -5 & -1 & 2 & 5 \\
$b_i$ & -5 & -4 & -1 & 2 & 5 \\
\hline
\end{tabular}
\end{center}
Clearly $a_i\leq b_i, \forall i$, but $\langle\svee_{i=1}^5 a_i\rangle=2$ while 
$\langle\svee_{i=1}^5 b_i\rangle=-4$.
\end{example}
On the contrary $\symsug_v$ and $\symsug^3_v$ are monotonic. Since in decision
making monotonicity is a mandatory property, $\symsug^2_v$ cannot be
used. Hence, we recommend the use of either $\symsug_v$ or $\symsug^3_v$. 
\end{itemize}

\section{Related works}
We are not aware of similar attempts to build algebraic structures on symmetric
ordered sets. We did an extension of this work in \cite{grbafo01}, where $L$
was restricted to $[-1,1]$, but the operator to be extended was any
t-conorm. The result is that an Abelian group can be built when the t-conorm is
strict, but a ring is not possible. There are connections of this result with
uninorms \cite{yary96}, and in a more general way, with partially ordered
groups and rings (see Fuchs \cite{fuc63}).

\medskip

Taking the viewpoint of decision making, as explained in the introduction, the
symmetric Sugeno integral may be used as the main ingredient of an ordinal
Cumulative Prospect Theory. Specifically, we aim at finding a representation of
preference over a set of alternatives or acts, when acts have as consequence
gains as well as losses, or put differently, when one can imagine the
\emph{symmetric} of a given act. There exist some works along this line, under
the name of \emph{signed orders}. We present briefly the concept of signed
order \cite{fis92}.

Let $X$ be a set of alternatives (possibly multidimensional), and $\succeq$ a
transitive complete relation on $X$. We consider a copy $X^*$ of $X$, whose
elements are denoted $x^*$ (\emph{reflection} of $x$). The relation $\succeq$
extended on $X\cup X^*$ is assumed to be self-reflecting:
\[
r\succeq s \Leftrightarrow s^*\succeq r^*.
\] 
$(X\cup X^*, \succeq)$ is called a \emph{self-reflecting signed
order}. Fishburn studies under which conditions a numerical representation $u$
of $\succeq$ can be found, i.e. such that $r\succeq s \Leftrightarrow u(r)\geq
u(s)$. It amounts that $u$ has to be skew symmetric, i.e. $u(r)+u(r^*)=0$. 

A similar attempt has been done by Suck \cite{suc84}, who proposed
\emph{compensatory structures}. $A$ being a set of alternatives, a compensatory
structure is a triplet $(A,L,K)$, where $L,K$ are binary relations, $L$ is
complete and transitive, $K$ is symmetric such that there exists $a'\in A$ for
each $a\in A$ so that $aKa'$, and if $aLb, aKa'$ and $bKb'$, then
$b'La'$. Clearly, $K$ is a reflection which reverses the order, like for
Fishburn's signed order. The representation theorem is indeed similar, and
includes the case of skew symmetry. 

Our approach and aim are however rather different, since we are dealing with
\emph{scores} or \emph{utilities}, on which we want to build some structure,
and not on the alternatives. It remains to embed the use of the symmetric
Sugeno integral in a decision making framework. We have already proposed such
an attempt for multicriteria decision making \cite{grdila01}.

\section*{Acknowledgement}
This work has benefited from many discussions, in particular with
D. Denneberg, J. Fodor, B. de Baets, R. Mesiar, J.L. Marichal, who are deeply
acknowledged.  We thank also the anonymous reviewers for their constructive
comments. 

\bibliographystyle{plain}
\bibliography{../BIB/fuzzy,../BIB/grabisch,../BIB/general}

\end{document}